 \documentclass[12pt,letterpaper]{article}
 \usepackage{amsmath, amsfonts, amssymb}
 \usepackage{epsfig}
 \usepackage{subfigure}
  \newcommand{\ket}[1]{\ensuremath{\left|{#1}\right>}}
  \newcommand{\bra}[1]{\ensuremath{\left<{#1}\right|}}

\begin{document}
 
 \title{Spin-dependent Bohm trajectories associated with an electronic
 transition in hydrogen}
 \author{C. Colijn and E. R. Vrscay \\ Department of Applied
 Mathematics \\ University of Waterloo \\
 Waterloo, Ontario, Canada N2L 3G1}
 \maketitle

 \begin{abstract}
 The Bohm causal theory of quantum mechanics with spin-dependence
 is used to determine electron
 trajectories when a hydrogen atom is subjected to (semi-classical)
 radiation.
 The transition between the $1s$ ground state and the
 $2p_0$ state is examined.  It is found that transitions can be identified
 along Bohm trajectories.
 The trajectories lie on invariant hyperboloid surfaces of revolution in
 ${\bf R}^3$. The energy along the trajectories is also discussed in
 relation to the hydrogen energy eigenvalues.
 \end{abstract}
 
 
 
 {\bf Short Title:}  Spin-dependent Bohm trajectories
 
 \newpage
 \section{Introduction}
 
 In the de Broglie-Bohm causal interpretation of quantum mechanics
 \cite{Bo52,BoHi,Ho93},
 as opposed to the standard ``Copenhagen'' interpretation,
 particles are endowed with well-defined trajectories ${\bf x}_i(t)$
 that are determined by the wavefunction $\psi({\bf x},t)$ of the
 quantum system being studied.
 A compatability with the statistical results of quantum
 mechanics is achieved by assigning an uncertainty in the
 initial conditions of the particles according to
 the probability density function $\rho(x,0) = |\psi(x,0)|^2$.
 
 The causal interpretation continues to receive attention as a way of
 addressing the question of the incompleteness of standard quantum
 mechanics (see, for example, \cite{AfSe,CuFiGo}).  Recently, however,
 causal trajectories have been playing a more significant role in
 practical calculations of chemical physics and quantum chemistry, for
 example (to name only a few references), quantum tunnelling dynamics
 \cite{Bi}, nonadiabatic transitions \cite{BuTu}, reactive scattering
 \cite{Wy}, dissociation dynamics \cite{WaDaHo} and hybrid
 classical/quantum schemes to study complex systems \cite{GiMeBe}.
 
 Bohm himself \cite{Bo52} introduced the idea of studying
 transitions in terms of the causal interpretation, examining
 the Franck-Hertz experiment and the photoelectric
 and Compton effects.  He attempted to show
 that the seemingly discontinuous and poorly defined
 transfers of energy and momentum in transitions could be accounted for
 in a continuous matter by means of the ``quantum potential''
 that arises in the causal formalism.
 More recently in this vein, Dewdney and Lam
 \cite{DeLa} studied transitions of (spinless) particles
 in a one-dimensional infinite square well potential.
 In this paper we employ the de Broglie-Bohm deterministic approach to study
 the problem of a $1s$-$2p$ electronic transition in
 hydrogen induced by
 an oscillating (semiclassical) electromagnetic field, taking
 the spin of the electron into account.
 
 It is instructive to review briefly
 the main ideas of the causal interpretation.
 First, the wavefunction $\psi$ for a particle is written in the form
 \begin{equation}
 \label{psi}
 \psi ( {\bf x},t ) = R ({\bf x},t) e^{i S({\bf x},t)/\hbar} ,
 \end{equation}
 where $R$ and $S$ are real-valued.  Substitution of Eq. \eqref{psi}
 into the time-dependent Schr\"{o}dinger equation,
 \begin{equation}
 \label{schrodinger}
 i \hbar \frac{\partial \psi}{\partial t} = - \frac{\hbar^2}{2m} \nabla^2 \psi + V \psi ,
 \end{equation}
 yields the following
 coupled equations in $R$ and $S$:
 \begin{equation}
 \label{requation}
 \frac{\partial R^2}{\partial t} + \nabla \cdot
 \left ( R^2 \frac{\nabla S}{m} \right ) = 0
 \end{equation}
 and
 \begin{equation}
 \label{sequation}
 \frac{\partial S}{\partial t} + \frac{(\nabla S)^2}{2m} + V + Q = 0,
 \end{equation}
 where
 \begin{equation}
 \label{quantumpotential}
 Q = - \frac{\hbar^2}{2m}\frac{\nabla^2 R}{R}
 \end{equation}
 is called the {\em quantum potential}.
 Eq. \eqref{requation} is the standard continuity equation of quantum mechanics.
 It can be viewed as governing the evolution of
 a compressible, irrotational fluid with density $\rho = \psi^*\psi = R^2$ and
 velocity ${\bf v} = \nabla S / m$ as, indeed, was done by
 Madelung \cite{Ma}.
 Eq. \eqref{sequation} has the same form as the classical
 Hamilton-Jacobi equation
 for a particle that moves under the influence of potentials
 $V$ and $Q$.
 Bohm's unique interpretation of these equations was that the particle has
 a well-defined trajectory defined by the
 {\em quantum equation of motion}
 \begin{equation}
 \label{momentum1}
 {\bf p} = m \dot{\bf x} = \nabla S .
 \end{equation}
 As stated earlier, compatability with standard quantum mechanics is achieved
 by viewing the initial conditions of trajectories as ``hidden variables''
 with associated uncertainties as described by the
 probability distribution $\rho({\bf x},0)$.
 
 Recall that the Schr\"{o}dinger current {\bf j} associated
 with the wavefunction $\psi$ is given by \cite{Sch}
 \begin{equation}
 \label{schrodingercurrent}
 {\bf j} = \frac{\hbar}{2mi} [ \psi^* \nabla \psi - \psi \nabla \psi^* ] .
 \end{equation}
 A comparison of Eqs. \eqref{momentum1} and \eqref{schrodingercurrent}
 shows that
 \begin{equation}
 \label{current1}
 {\bf j} = \frac{1}{m} \rho {\bf p}.
 \end{equation}
 
 The momentum defined in Eq. \eqref{momentum1} is not unique
 in generating the same statistical predictions as quantum mechanics.
 Holland \cite{Ho99} has shown that Eqs. \eqref{momentum1} and \eqref{current1}
 apply only to spinless particles.  For particles with spin,
 an additional term is necessary in order that the Schr\"{o}dinger
 equation of motion be consistent with a relativistic formulation.
 The condition of Lorentz invariance implies that the momentum of
 a particle with spin {\bf s}, even in the non-relativistic limit, must be given by
 \begin{equation}
 \label{momentum2}
 {\bf p} = \nabla S + \nabla \log \rho \times {\bf s} .
 \end{equation}
 The associated current
 \begin{equation}
 \label{current2}
 {\bf j} = \frac{1}{m} \rho \nabla S + \frac{1}{m} \nabla \rho \times {\bf s} ,
 \end{equation}
 has been referred to as the {\em Pauli current}, the nonrelativistic
 limit of the {\em Dirac current}, as opposed to Eq. \eqref{current1} which
 is the nonrelativistic limit of the {\em Gordon current} \cite{GuHe,He75,He79}.
 Consistency with Dirac theory requires that the Schr\"{o}dinger equation
 be regarded as describing an electron in an eigenstate of spin \cite{GuHe}.
 As regards the causal interpretation,
 the spin-dependent term was also discussed in \cite{BoHi} but only
  for the Pauli equation and not the Schr\"{o}dinger equation.
 
 In the case of a hydrogen atom,
 the momentum equation \eqref{momentum1} predicts that an electron
 in any real eigenstate will be stationary since $\nabla S = 0$.
 This counterintuitive result no longer applies when Eq. \eqref{momentum2}
 is used.  For example, consider
 an electron in the $1s$ ground eigenstate with wavefunction
 \begin{equation}
 \label{psi_1s}
 \psi_{100} = \frac{1}{\sqrt{\pi a^3} } e^{-r/a} ,
 \end{equation}
 where $a = \hbar^2/(me^2)$ is the Bohr radius.
 Also assume that the electron is in a definite spin eigenstate
 so that its spin vector is given by
 ${\bf s} = \frac{\hbar}{2}{\bf k}$.
 Then the existence of the term
 $\nabla \log \rho \times {\bf s}$ in Eq. \eqref{momentum2} implies
 that the electron's polar coordinates $r$ and $\theta$ are constant
 and that the angle $\phi$ evolves in time as \cite{Ho93,CoVr}
 \begin{equation}
 \label{phi_1s}
 \frac{d \phi}{dt} = \frac{\hbar}{mar}.
 \end{equation}
 Therefore, the electron revolves about the $z$-axis at
 a constant angular velocity.
 
 We also state, for future reference,
 the result for an electron in the (real) $2p_0$ eigenstate
 \begin{equation}
 \label{psi_210}
 \psi_{210}=\frac{1}{\sqrt{32a^5}}
 re^{-r/2a}\cos\theta,
 \end{equation}
 again with spin vector ${\bf s} = \frac{\hbar}{2}{\bf k}$:
 The polar coordinates $r$ and $\theta$
 are again constant and the angle $\phi$ evolves as
 \begin{equation}
 \label{phi_2p0}
 \frac{d \phi}{dt} = \frac{\hbar}{2mar}.
 \end{equation}
 Note that the angular velocity
 is one-half that
 of a $1s$ ground state electron.
 In \cite{CoVr} we examined the trajectories of electrons in a number of
 hydrogenic eigenstates.  From Eq. \eqref{momentum2}, these trajectories must
 lie on level curves of both $|\psi|$ and $z$ and revolve
 about the $z$-axis with constant angular velocity.
 
 In this paper we examine solutions of the equation of motion \eqref{momentum2}
 for an electron with spin vector
 ${\bf s} = \frac{\hbar}{2}{\bf k}$ (the ``$\alpha$'' or ``spin up'' state)
 as it undergoes
 a transition from the $1s$ to $2p_0$ state in hydrogen due to the presence of
 an oscillating electric field.
 We may assume that the electron has constant spin vector since the
 hamiltonian describing the atom in the field (see next section) is spin-independent.
 The wavefunction of the electron $\Psi({\bf x},{\bf s},t)$
 may then be written as the tensor product $\psi({\bf x},t)\zeta({\bf s})$
 where $\zeta({\bf s})$ is assumed to be an eigenfunction of the commuting spin operators
 $\hat{S}^2$ and $\hat{S}_z$, with $\hat{S}^2 \zeta = \frac{3\hbar}{4} \zeta$
 and $\hat{S}_z \zeta = \frac{\hbar}{2} \zeta$.
 As such, the remainder of our discussion may simply be focussed on the
 evolution of the spatial portion of the wavefunction, $\psi({\bf x},t)$
 according to Eq. \eqref{schrodinger}.
 
 As in the case of the examples listed above,
 the momentum term $\nabla \log \rho \times {\bf s}$ will be responsible for
 the revolution of the electron about the $z$-axis.
 This is accompanied by a complicated motion
 over a hyperboloid surface of revolution that is determined
 from the functional forms of the $1s$ and $2p_0$ wavefunctions.
 Moreover, the course of the transition from the
 ground state to the excited state can be characterized
 by looking at the energy of the electron along the trajectory
 and the angular velocity of the revolution about the $z$-axis.
 The energy and $\phi$-angular velocity
 of the electron evolve from
 $1s$ ground state values to
 $2p$ excited state values.
 
 In \cite{CoVr}, as a precursor to this study,
 we examined the trajectories dictated
 by Eq. \eqref{momentum2} for an electron with
 spin vector ${\bf s} = \frac{\hbar}{2}{\bf k}$ and spatial wavefunction
 that begins as a linear combination of $1s$ and $2p$
 hydrogenic eigenfunctions:
 \begin{equation}
 \label{modelproblem0}
 \psi({\bf x},0) = c_1 \psi_{100}({\bf x}) +
                   c_2 \psi_{210}({\bf x}) ,
 \end{equation}
 where $|c_1|^2 + |c_2|^2 = 1$.  The time evolution of this wavefunction
 under the hydrogen atom hamiltonian is
 \begin{equation}
 \label{modelproblemt}
 \psi({\bf x},t) = c_1 \psi_{100}({\bf x}) e^{-iE_1t/\hbar} +
                   c_2 \psi_{210}({\bf x}) e^{-iE_2t/\hbar} .
 \end{equation}
 Many of the qualitative features of the $1s$-$2p_0$ transition problem
 studied below are captured by this model, most notably the
 invariant hyperboloid surfaces of revolution on which trajectories
 lie.  As expected, however, the more detailed time evolution of the
 electron trajectories over these surfaces due to the oscillating field is missing.
 
 \section{Solution of the transition problem}
 
 The hamiltonian used to describe this transition will have the
 form $\hat H = \hat{H}_0 + \hat{H}^\prime$, where $\hat{H}_0$
 is the hydrogen atom hamiltonian and $\hat{H}^\prime$ represents
 an oscillating electric field ${\bf E} = E_0 \cos \omega t {\bf k}$.
 It can be shown \cite{Gr} that if $\omega$ is chosen
 to be sufficiently close to the
 $1s$-$2p$ transition frequency
 \begin{equation}
 \label{omega0}
 \omega_0 = \frac{E_2 - E_1}{\hbar} ,
 \end{equation}
 so that $\omega_0 - \omega \ll \omega_0 + \omega$, then the
 hamiltonian representing the semiclassical radiation,
 $\hat{H}'=q z E_0 \cos \omega t$, is well approximated by
 \begin{equation}
 \label{hamiltonianp}
 \hat{H}^\prime = -\frac{1}{2} q z E_0 e^{-i \omega t},
 \end{equation}
 since the term $i (\omega + \omega_0)^{-1} \sin \omega t$
 is negligible.
 Here, $q$ denotes the electric charge and
 $\omega_0~\approx~1.549~\times~10^{16}~\text{s}^{-1}$.
 This approach allows the equations for the wavefunction coefficients
 to be solved exactly so that perturbation methods need not be employed.
 The closeness of $\omega$ to $\omega_0$ also allows the transition
 probability to approach unity at various times rather than remaining
 small for all times.
 
 The probability of transition between two states $\ket{\psi_1}$
 and $\ket{\psi_2}$ is related to the matrix element
 $\bra{\psi_2}\hat{H}'\ket{\psi_1}$. In the case of the hydrogen atom the
 only nonvanishing matrix element
 is between the ground state $\psi_{100}$ and the $2p_0$ state $\psi_{210}$:
 \begin{eqnarray}
 \bra{\psi_{100}}\hat{H}'\ket{\psi_{210}} & = &
           -\bra{\psi_{100}}qE_0
            r \cos \theta\ket{\psi_{210}}\frac{1}{2}e^{-i\omega t} \nonumber \\
      & = & -\frac{64\sqrt{2}}{243}a q E_0 e^{-i\omega t} .
 \end{eqnarray}
 The time-dependent wavefunction can be written as a linear combination of
 $\ket{\psi_{100}}$ and $\ket{\psi_{210}}$:
 \begin{equation}
 \label{psicombo}
 \psi(t)=c_a(t)\psi_{100}e^{-i E_1 t/\hbar} +
 c_b(t)\psi_{210}e^{-i E_2 t/\hbar}.
 \end{equation}
 Substitution into the Schr\"{o}dinger equation
 yields the following equations
 for $c_a(t)$ and $c_b(t)$:
 
 \begin{equation}
 \begin{split}\label{cacbde}
 \dot{c_a} &= -\frac{i}{\hbar}\frac{V_{12}}{2}e^{-i(\omega_0-\omega)t}
 c_b, \\
 \dot{c_b} &= -\frac{i}{\hbar}\frac{V_{12}}{2}e^{i(\omega_0-\omega)t}
 c_a,
 \end{split}
 \end{equation}
 where
 \begin{displaymath}
 V_{12} = -\frac{128 \sqrt{2}}{243}aq E_0.
 \end{displaymath}
 These DEs
 can be solved exactly to give
 \begin{equation}\label{cacb}
 \begin{split}
 c_a(t)&=\frac{\sigma + \Omega}{2\sigma}e^{\frac{1}{2}i(\Omega-\sigma)t}
 + \frac{\sigma - \Omega}{2\sigma}e^{\frac{1}{2}i(\Omega+\sigma)t}, \\
 c_b(t)&=\frac{\nu}{2\sigma}e^{\frac{1}{2}i(\Omega-\sigma)t}-
  \frac{\nu}{2\sigma}e^{\frac{1}{2}i(\Omega+\sigma)t},
 \end{split}
 \end{equation}
 where
 \begin{equation}
 \begin{split} \Omega &= \omega_0 -\omega, \\
  \nu &= \frac{V_{12}}{\hbar}, \\
  \sigma &= \sqrt{\Omega^2+\nu^2}.
 \end{split}
 \end{equation}
 
 The wavefunction $\psi({\bf x}, t)$ may now be written explicitly as
 \begin{equation}
 \label{tdpsi}
 \psi({\bf x},t)
  = \frac{1}{\sqrt{\pi a^3}}c_a(t)e^{-r/a}e^{-i E_1 t/\hbar}
 + \frac{1}{\sqrt{32\pi a^5}}c_b(t)r e^{-r/2a}\cos \theta e^{-i E_2
 t/\hbar}.
 \end{equation}
 To compute the momentum according to Eq. \eqref{momentum2}, note that
 the wave function, and hence
 first term $\nabla S$, has only $\hat{r}$ and $\hat{\theta}$
 components.   Since we are assuming a constant spin vector ${\bf
 s}=\frac{\hbar}{2}\hat{k}$, it follows that the vector
 $\nabla \log\rho \times {\bf s}$ points in the
 $\hat{\phi}$ direction.
 
 Calculating $\nabla S$ from Eq. \eqref{tdpsi} yields
 \begin{equation}\begin{split}
 \label{sdes}
 p_{\hat{r}} &= \frac{\hbar \nu\beta}{2\sigma}\frac{ \cos\theta
 e^{-3r/2a}(1+\frac{r}{2a})T(t)}{D(r,\theta,t)}, \\
 p_{\hat{\theta}} &= \frac{\hbar \nu\beta}{2\sigma}
 \frac{ \sin\theta e^{-3r/2a}T(t)}{D(r,\theta,t)},\end{split}
 \end{equation}
 where $\beta = 4 \sqrt{2} a$ is the ratio of the normalizing factors of
 the two wavefunctions,
 \begin{equation}\label{timefunc}
 T(t)=-\cos \omega_0 t \sin \sigma t -
  \frac{\Omega}{\sigma}\sin \omega_0 t +\frac{\Omega}{\sigma} \cos
 \sigma t \sin \omega_0 t \end{equation}
 and
 \begin{multline}
 D(r,\theta,t)=e^{-2r/a} \frac{1}{2\sigma^2} (\sigma^2+ \Omega^2 +\nu^2\cos \sigma t)+
 \beta^2 r^2 e^{-r/a} \cos^2\theta
 \frac{\nu^2}{2\sigma^2}  (1-\cos \sigma t) \\
 + \frac{\nu}{\sigma}\beta r e^{-\frac{3r}{2a}}\cos
 \theta (\frac{\Omega}{\sigma}
 \cos \omega_0 t -\frac{\Omega}{\sigma}\cos \sigma t \cos \omega_0 t
 - \sin \omega_0 t \sin \sigma t).
 \end{multline}
 The denominator $D(r,\theta,t)$ in the above expressions is proportional to
 $|\psi({\bf x},t)|^2$.
 
 The second term in Eq. \eqref{momentum2}, $\nabla \log\rho \times {\bf s}$,
 can be computed in the (right-handed) spherical polar coordinate system
 \begin{equation}\label{polarcross}
  \mathbf{A}\times\mathbf{B}=
 \begin{vmatrix} \hat{\theta} & \hat{\phi}& \hat{r}\\
 				A_{\theta} & A_{\phi} & A_{r} \\
 				B_{\theta} & B_{\phi} & B_{r} \end{vmatrix} .
 \end{equation}
 Using the simplification
 \begin{displaymath}
 \nabla
 \log\rho=2\text{Re}\left(\frac{(\nabla\psi)\psi^*}{\psi^*\psi}\right),
 \end{displaymath}
 we find the $\hat{\phi}$ component of the momentum to be
 \begin{equation}\label{phieq}
 p_{\hat{\phi}}=\frac{\hbar\beta}{D}(-\chi_r \sin\theta -
 \chi_{\theta}\cos\theta),
 \end{equation}
 where
 \begin{align*}
  \chi_r &= \frac{1}{\beta a} |c_a|^2
 e^{-2r/a}+
 \beta |c_b|^2 \cos^2\theta e^{-r/a}r(1-\frac{r}{2a})+
 \cos\theta e^{-3r/2a}(1-\frac{3r}{2a})T'(t), \\
  \chi_{\theta}&=-\beta |c_b|^2
 e^{-r/a}\sin\theta
 \cos\theta r - e^{-3r/2a}\sin\theta \; T'(t)
 \end{align*}
 and
 \begin{equation}\label{Tprime}
  T'(t)=\frac{\nu}{2\sigma}(\frac{\Omega}{\sigma}\cos\omega_0
  t-\frac{\Omega}{\sigma} \cos\sigma t \cos\omega_0 t -
 \sin\sigma t\sin\omega_o t). \end{equation}
 In summary, the three components of the momentum are given by
 Eqs. \eqref{sdes} and \eqref{phieq}.
 It is worth noting  that the spin-dependent momentum term
 $\nabla \log \rho \times {\bf s}$ in Eq. \eqref{momentum2} is
 responsible for the $\phi$-momentum $p_{\hat{\phi}}$.
 
 It is useful to rescale these equations by defining the
 following dimensionless variables:
 \begin{equation}
 \label{dimensionless}
 \xi=\frac{r}{a}, ~~~~~ \tau = \omega_0 t .
 \end{equation}
 In these variables, Eqs. \eqref{sdes} and \eqref{phieq} give rise to the
 following system of differential equations in $\xi$, $\theta$, and
 $\phi$ as functions of $\tau$:
 \begin{equation}\label{scaleddes}
 \begin{split}
 \frac{d\xi}{d\tau} &= \frac{\nu}{3\sqrt{2} \sigma}
 \left(\cos\theta e^{-3\xi/2}(1+\frac{\xi}{2})\right)
 \frac{\tilde{T}(\tau)}{D} \\
 \frac{d\theta}{d\tau}&= \frac{\nu}{3\sqrt{2} \sigma}
 \left(\sin\theta \frac{e^{-3\xi/2}}{\xi}\right)
 \frac{\tilde{T}(\tau)}{D}\\
 \frac{d\phi}{d\tau} &= -\frac{\nu}{3\sqrt{2} \sigma\xi D}
 (\chi_r+\chi_{\theta}\cot\theta),
 \end{split}
 \end{equation}
 where we have used the following relations:
 \begin{displaymath}
 \frac{d\phi}{d\tau}= \frac{p_{\phi}}{ma\omega_0\xi\sin\theta}
 \end{displaymath}
 and
 \begin{displaymath}
 \frac{\hbar}{ma\omega_0}=\frac{8}{3}a.
 \end{displaymath}
 
 Some important qualitative features of the solutions to these
 DEs may be extracted.
 First note that the $\xi$ and $\theta$ DEs are independent of $\phi$.
 >From these two equations, we have
 \begin{equation}
 \label{xithetade}
 \frac{d\xi}{d\theta} = - \xi\left(1+\frac{\xi}{2}\right)\cot\theta .
 \end{equation}
 This DE is easily solved to give  \cite{CoVr}
 \begin{equation}
 \label{hyperbola}
 \xi = \frac{2}{A \sin \theta - 1}, ~~~~~
 A = \frac{2 + \xi_0}{\xi_0 \sin \theta_0} > 1 ,
 \end{equation}
 where $\xi_0 = \xi(0)$ and $\theta_0 = \theta(0)$. This relation defines a
 family of hyperbolae in vertical planes that contain the $z$-axis.
 i.e., $\phi = \phi_0$, a constant.
 (Note that the right hand side of Eq. \eqref{xithetade} is determined
 by the functional forms of the $1s$ and $2p_0$ wavefunctions.  Other
 allowable pairs of wavefunctions will yield different types of curves.)
 The solutions of Eq. \eqref{scaleddes}
 therefore lie on surfaces that are obtained by rotating these hyperbolae
 about the $z$-axis (see \cite{CoVr} for a more detailed discussion).
 However, the time-dependent behaviour of the trajectories lying on
 these invariant sets must be determined numerically.
 
 \section{Numerical results}
 
 In choosing the parameters for the numerical integration, there are
 various factors that must be taken into account. First, recall that in
 order to use the hamiltonian $\hat{H}'= -\frac{1}{2} q z E_0 e^{-i\omega
 t}$, we require that the perturbing frequency $\omega$ be close to the
 transition frequency $\omega_0$ so that
 $\Omega = \omega_0 - \omega \ll \omega_0 + \omega$.
 Therefore, we cannot
 allow $\Omega$ to be too large, i.e., $\Omega \leqslant O(10^{13})$.
 
 If we wish to be fairly confident that a transition will occur,
 it is also necessary that the coefficient $c_b$ become large in
 magnitude at some time, i.e., roughly unity.  Recall from Eq.
 \eqref{cacb} that $c_b \propto \nu/2\sigma$, and $\nu \propto
 E_0$. Also, the derivatives in  Eq. \eqref{scaleddes}  are proportional to
 $\nu/2\sigma$, and it is desirable for numerical integration that they
 evaluate to order 1. Therefore, $|\nu|$ should be significant
 compared to $2\sigma$. Now, recall that
 $\sigma = \sqrt{\Omega^2+\nu^2}$ so that
 \begin{displaymath}
 \frac{|\nu|}{2 \sigma}= \frac{1}{2} \left [ \frac{\Omega^2}{\nu^2}+1 \right ]^{-1/2}.
 \end{displaymath}
 Therefore we require that $|\nu|$ not be too large. Because $|\nu|$ is
 proportional to the field strength, we are free to choose a small value.
 
 Another factor to consider is that there are two angular frequencies in the
 problem, namely $\sigma$ and $\omega_0$ from
 Eqs. \eqref{timefunc} and \eqref{Tprime}.
 For numerical stability it is best
 if these are within several orders of magnitude of each
 other. Therefore, $\sigma$ should not be too small in comparison with
 $\omega_0$.
 It is also desirable that $c_b$ become significant in a
 reasonable time. From Eq. \eqref{cacb}, we find that
 \begin{equation}
 \label{cb}
 |c_b(t)|^2 = \left ( \frac{\nu}{\sigma} \right )^2 \sin^2 \frac{\sigma}{2} t .
 \end{equation}
 Therefore, having $\sigma$
 appropriately scaled will mean that there is a high probability of
 seeing a transition within a reasonable time.
 
 The above considerations imply that:
 \begin{enumerate}
 \item
 $|\nu|$ cannot be too large
 because $[\Omega^2/\nu^2+1]^{-1/2}$ must be O(1),
 \item
 $\sigma =\sqrt{\Omega^2+\nu^2}$ cannot be more than several orders of
 magnitude smaller than $\omega_o$ and
 \item
 we cannot
 increase $\sigma$ by increasing $\Omega$, because
 we require that $\Omega \leqslant O(10^{13})$.
 \end{enumerate}
 With these points in mind, the
 parameters have been chosen as follows:
 \begin{equation}
 \begin{split}
 E_0 &= 8.8 \times 10^{7} ~ \text{V/m},  \\
 \Omega &= 1.55 \times 10^{12} \text{s}^{-1},
 \end{split}
 \end{equation}
 so that
 \begin{equation}
 \begin{split}
 \nu&= - 5.1\times 10^{12},\\
  \sigma&=5.32 \times 10^{12}.
 \end{split}
 \end{equation}

 \begin{figure}
 \begin{center}
 \epsfig{file=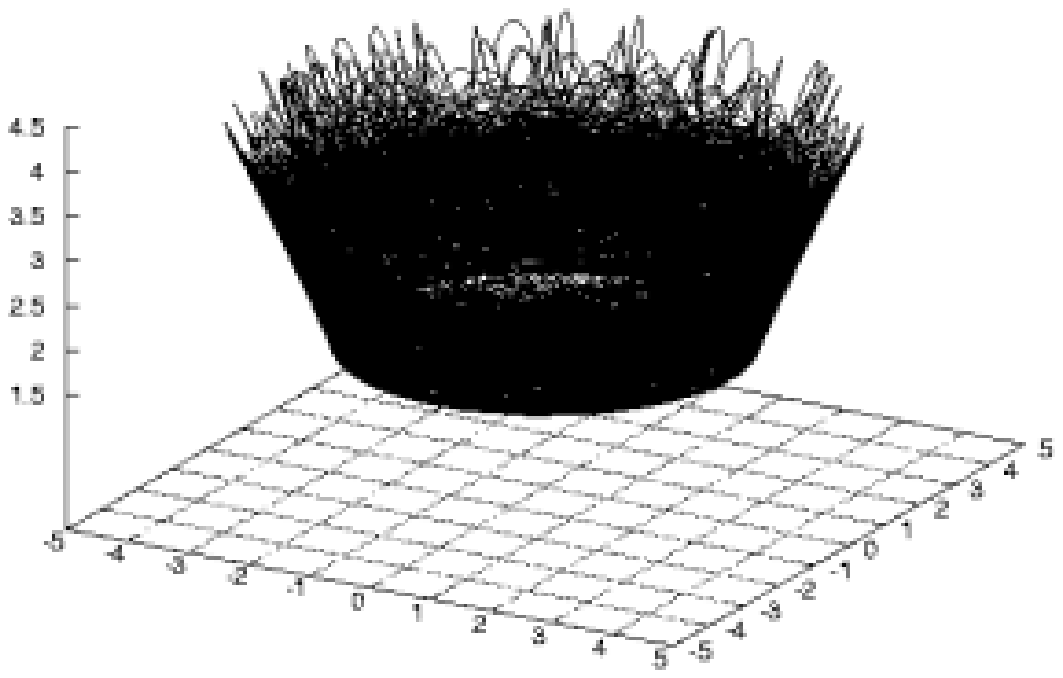,width=8cm,angle=270}
 \vspace{0.5cm}
 \end{center}
 \caption{Trajectory of electron in ${\bf R}^3$ during 
 $1s$-$2p_0$ transition: $\xi(0)=4$, $\theta(0)=1$}\label{fig:traja}
 \end{figure}
 \begin{figure}
 \begin{center}
 \epsfig{file=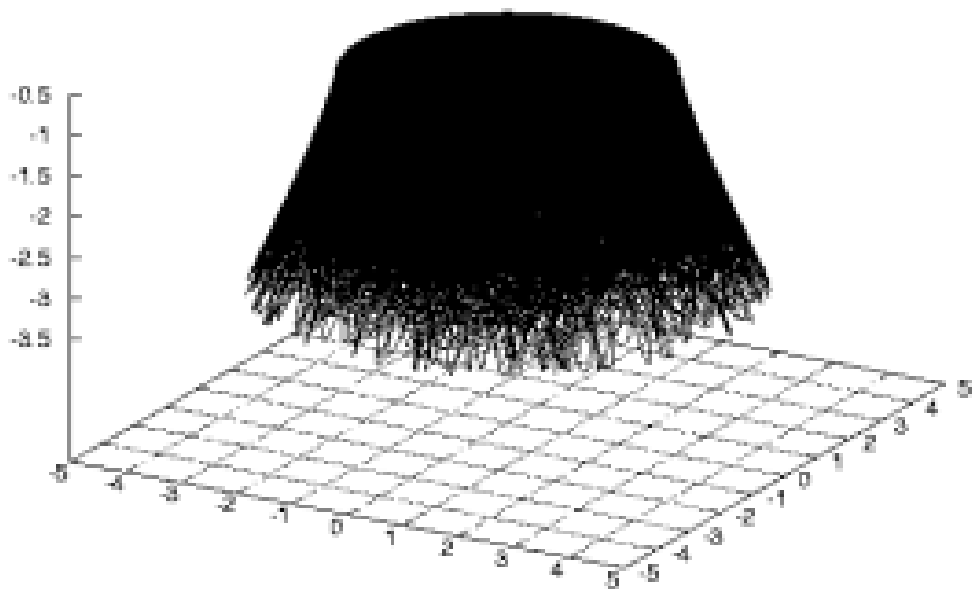,width=8cm,angle=270}
 \vspace{0.5cm}
 \end{center}
 \caption{Trajectory of electron in ${\bf R}^3$ during
 $1s$-$2p_0$ transition: $\xi(0)=3.2$, $\theta(0)=2$}\label{fig:trajb}
 \end{figure}
 \begin{figure}
 \begin{center}
 \mbox{\subfigure[$\tau=0-1469$]
 {\epsfig{file=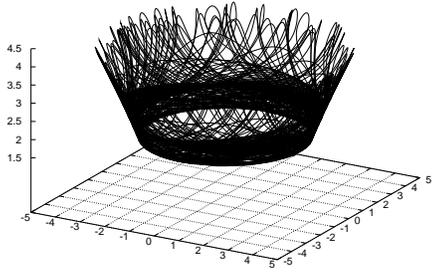,width=0.3\textwidth,angle=270}}\quad
 \subfigure[$\tau=1469-2992$]
 {\epsfig{file=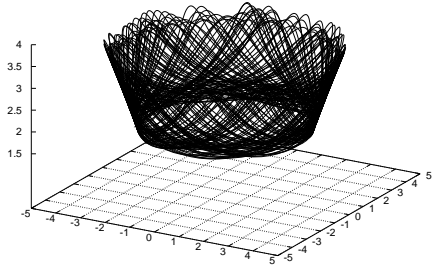,width=0.3\textwidth,angle=270}}}
 \mbox{\subfigure[$\tau=2992-4844$]
 {\epsfig{file=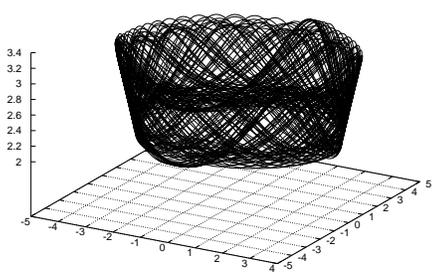,width=0.3\textwidth,angle=270}}\quad
 \subfigure[$\tau=4844-7196$]
 {\epsfig{file=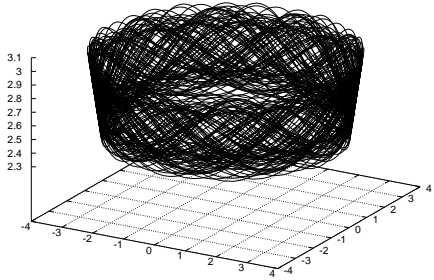,width=0.3\textwidth,angle=270}}}
 \subfigure[$\tau=7196-10000$]
 {\epsfig{file=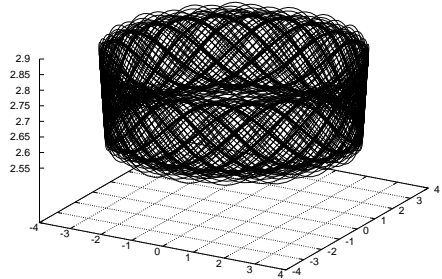,width=0.3\textwidth,angle=270}}
 \caption{Trajectory of electron in ${\bf R}^3$ during $1s$-$2p_0$ transition,
 split over five time intervals:  $\xi(0)=4$, $\theta(0)=1$}
 \label{fig:asplit}
 \end{center}
 \end{figure}
 \begin{figure}
 \begin{center}
 \mbox{\subfigure[$\tau=0-1469$]
 {\epsfig{file=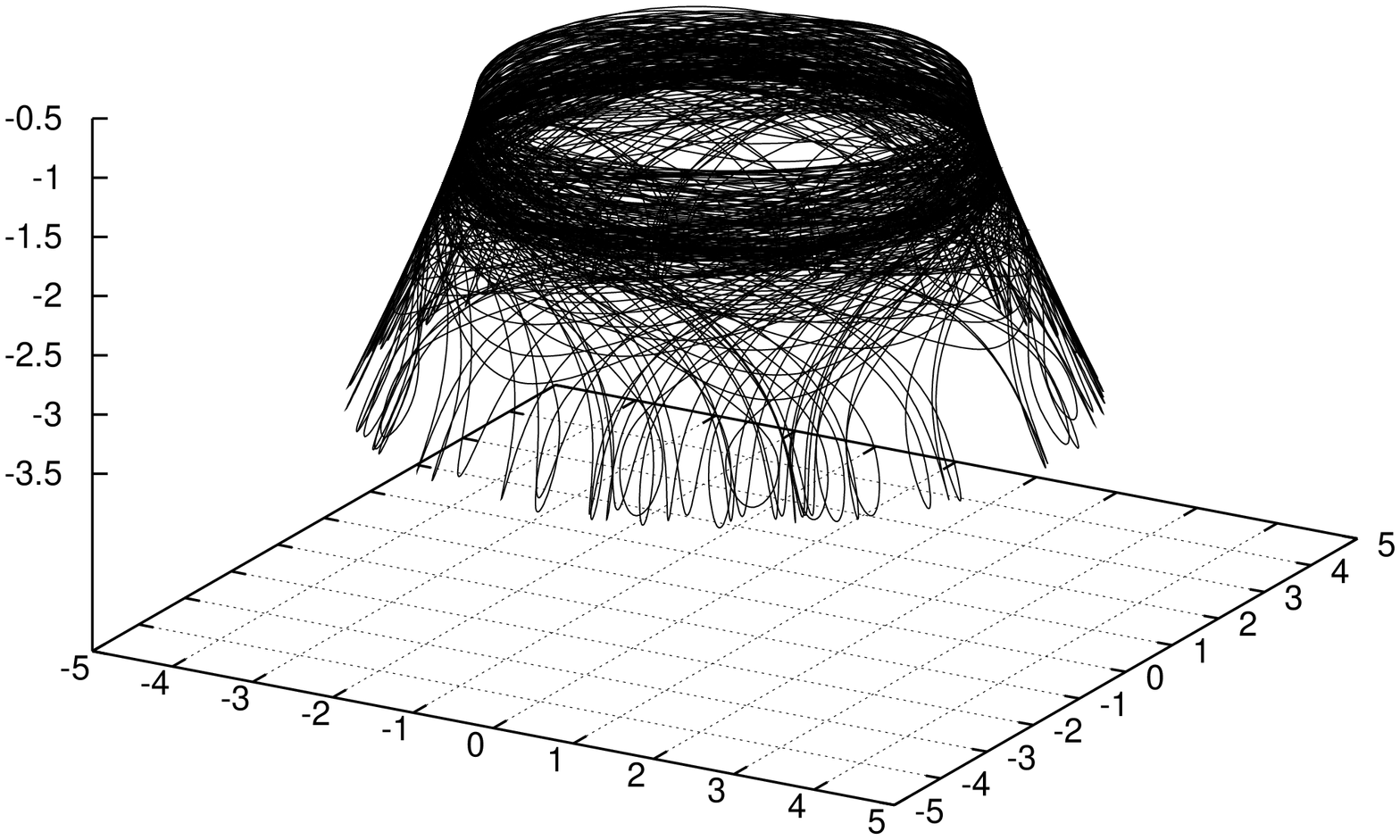,width=0.3\textwidth,angle=270}}\quad
 \subfigure[$\tau=1469-2992$]
 {\epsfig{file=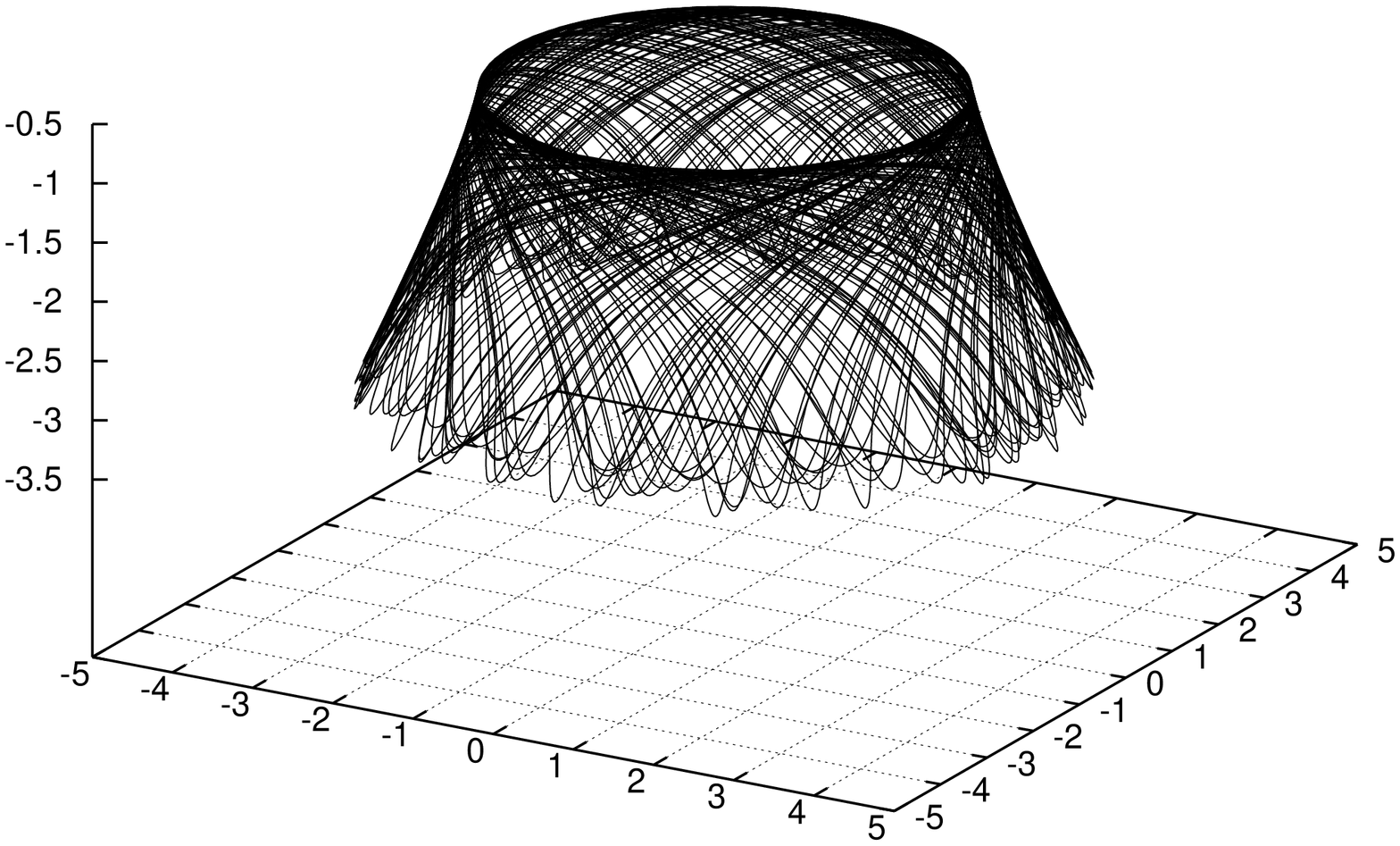,width=0.3\textwidth,angle=270}}}
 \mbox{\subfigure[$\tau=2992-4844$]
 {\epsfig{file=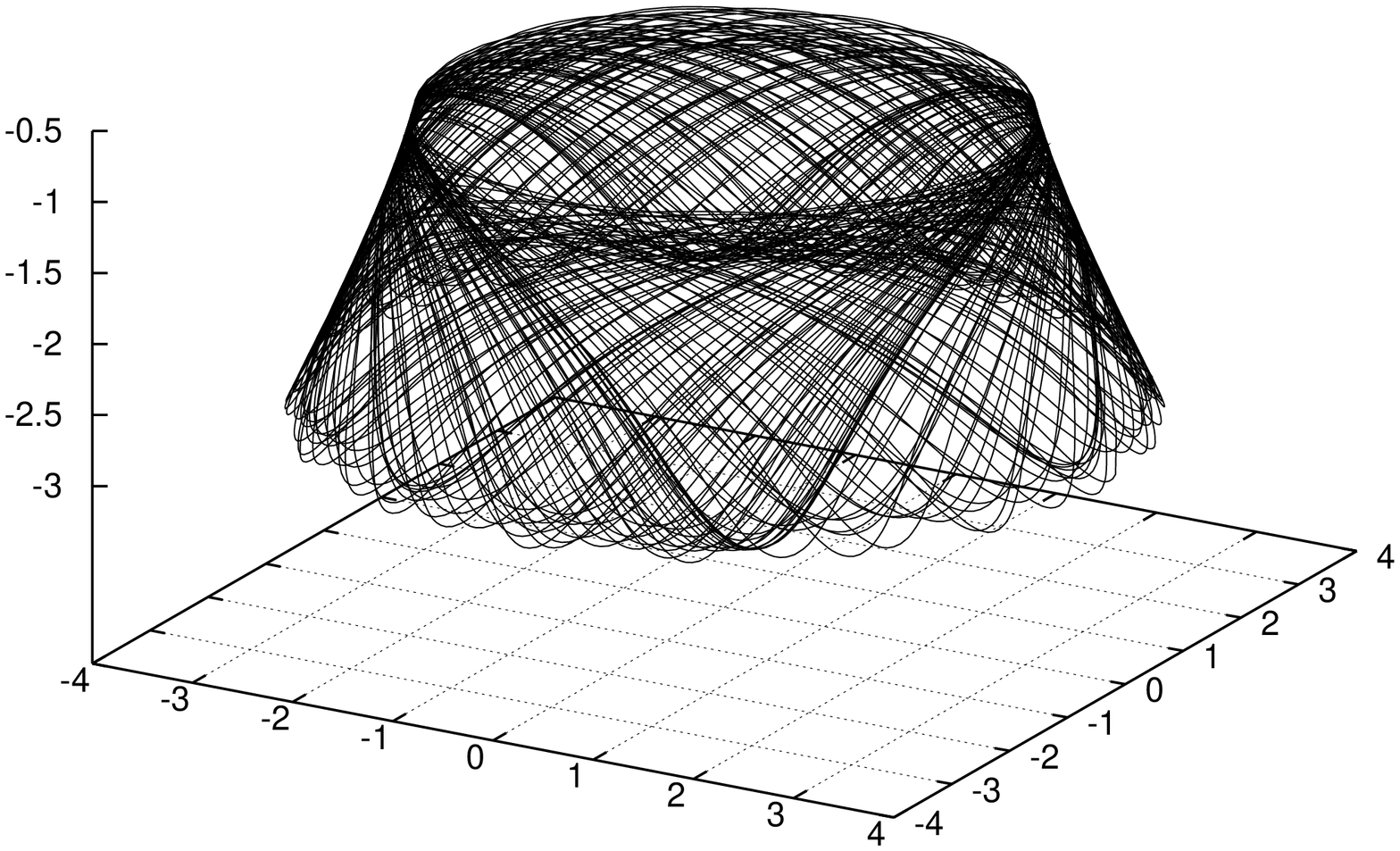,width=0.3\textwidth,angle=270}}\quad
 \subfigure[$\tau=4844-7196$]
 {\epsfig{file=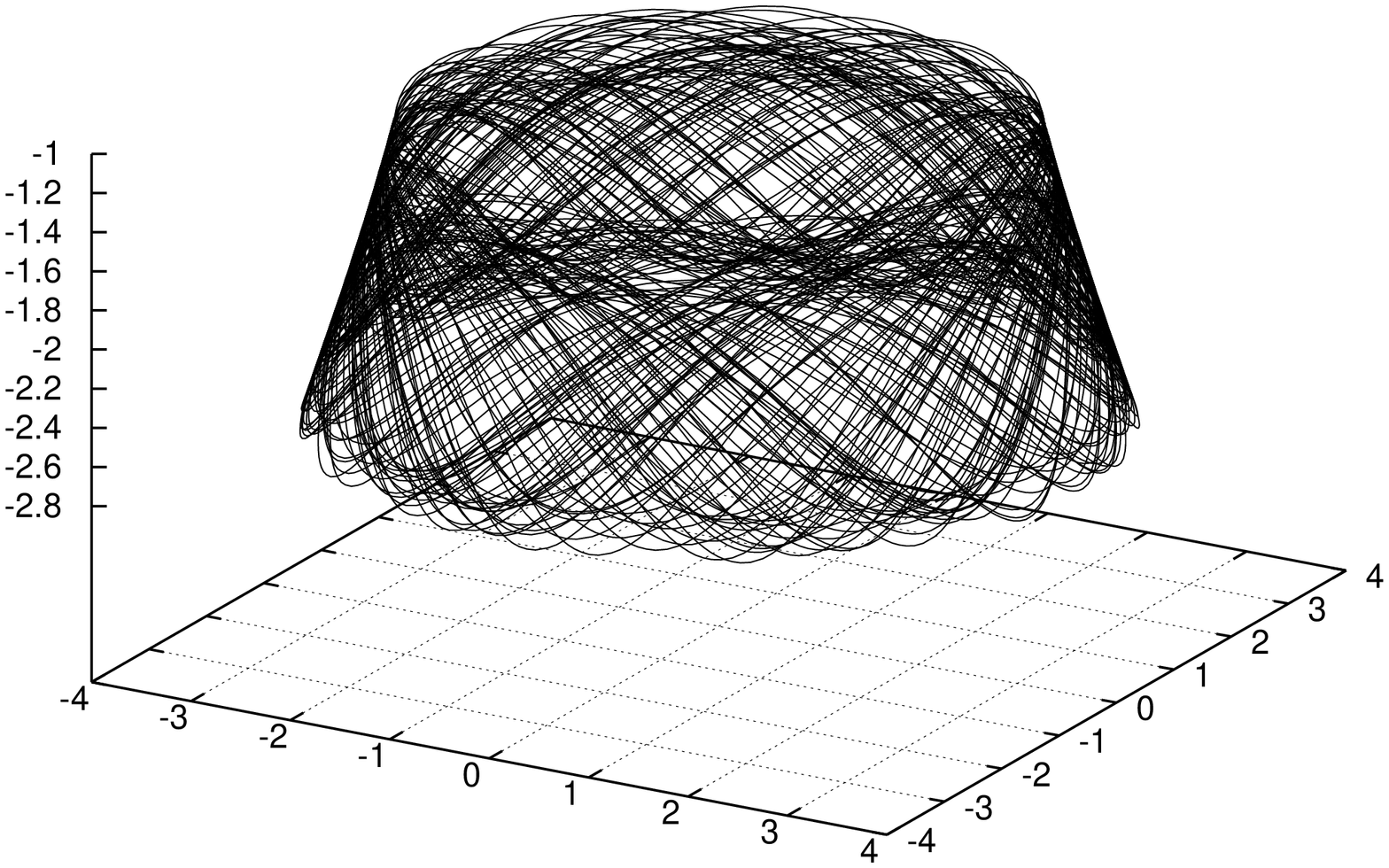,width=0.3\textwidth,angle=270}}}
 \subfigure[$\tau=7196-10000$]
 {\epsfig{file=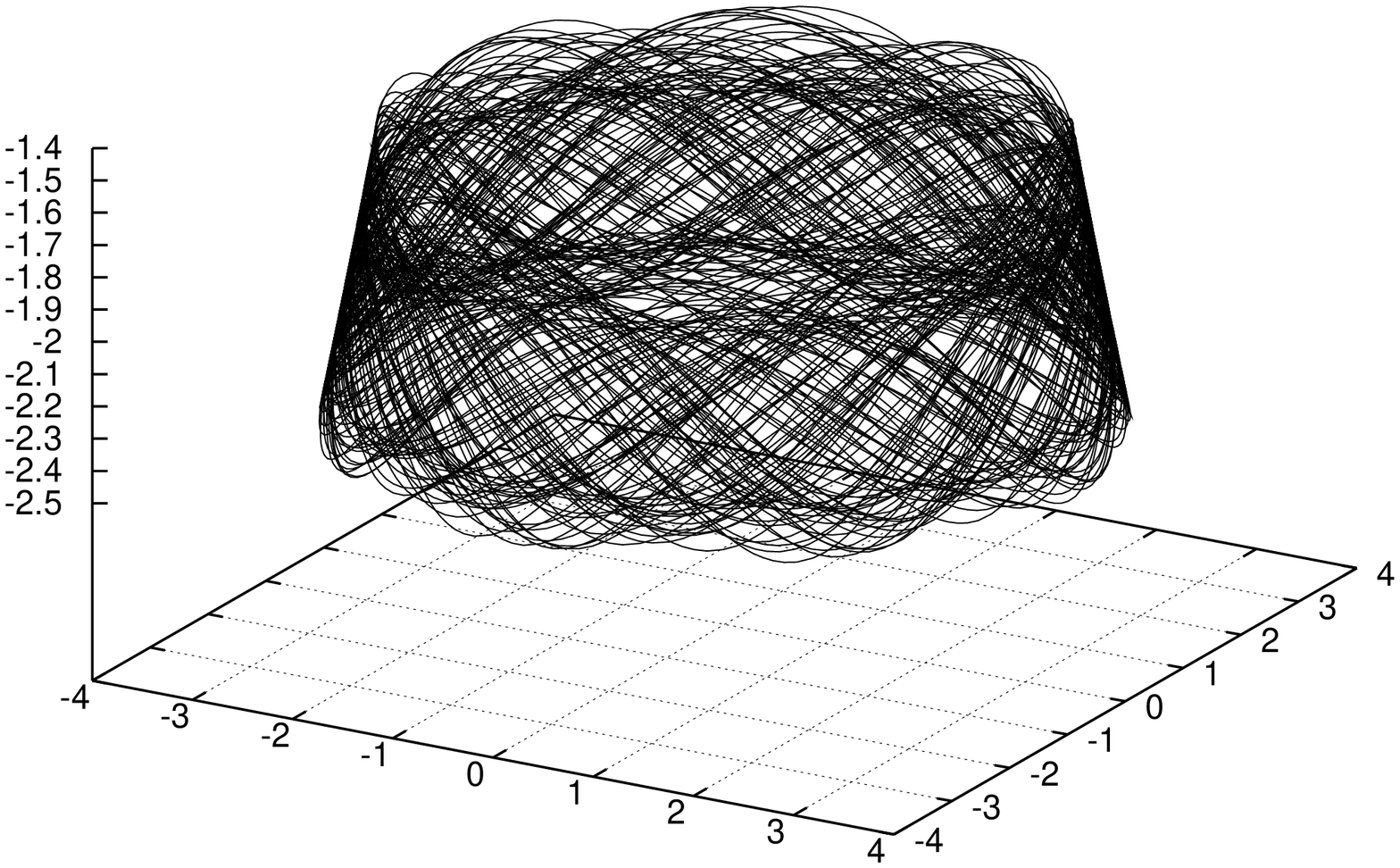,width=0.3\textwidth,angle=270}}
 \caption{Trajectory of electron in ${\bf R}^3$ during
 $1s$-$2p_0$ trajectory, split over five time intervals:  
 $\xi(0)=3.2$, $\theta(0)=2$}
 \label{fig:bsplit}
 \end{center}
 \end{figure}
 The results of numerical integration for two choices of initial conditions
 are shown in Figs.~\eqref{fig:traja} and
 \eqref{fig:trajb}. Figs.~\eqref{fig:asplit} and \eqref{fig:bsplit} show the
 same trajectories split over five consecutive time intervals.
 In these latter plots, the hyperboloid surfaces of revolution on which
 the trajectories lie, described at the end of the previous section, are discernable.
 
 In standard semiclassical treatments of this problem (see, for example,
 \cite{Sch}, pp. 282-285), the maximum probability of
 transition occurs when $\sin^2 \frac{1}{2}\Omega t =1$, or $\Omega t =
 (2k+1)\pi$ for $k=0,1,2,\dots$. In our formulation, from Eq. \eqref{cb}
 the maximum probability of transition occurs when
 $\sin^2 \frac{1}{2}\sigma t = 1$, or
 $\sigma t = (2k+1)\pi$ for $k=0,1,2,\dots$.
 In other words, the
 dependence on $\Omega$ is replaced with one on
 $\sigma = \sqrt{\Omega^2+\nu^2}$. This is
 due to our choice of hamiltonian and the fact that the equations in
 \eqref{cacb} result from an exact integration of Eq. \eqref{cacbde} rather than
 a perturbation approach.
 >From the above, the first occurrence of the maximum probability
 of transition in our problem will occur at $t = \pi/\sigma$ or,
 in dimensionless time, $\tau = \pi \omega_0 / \sigma$.
 >From the numerical values chosen for the parameters above, this corresponds
 to $\tau \approx 9000$.
 
 The causal interpretation allows us to examine the process in more
 detail by looking at the angular velocity $\dot \phi$ exhibited by
 the trajectories.
 At time $t = \tau = 0$,
 each trajectory begins at the $1s$ ground state wavefunction, viz., $c_a(0)=1$ and
 $c_b(0)=0$, and we expect the electron to revolve about the $z$-axis
 with angular frequency given by Eq. \eqref{phi_1s}.
 When the wavefunction is approximately
 equal to the $2p_0$ eigenstate, viz., $c_a \approx 0$ and $c_b \approx 1$,
 then we expect $\dot \phi$ to be given roughly by Eq. \eqref{phi_2p0}.
 
 Since the computations were performed in scaled variables $\xi$ and $\tau$,
 cf. Eq. \eqref{dimensionless}, we must first rescale the $1s$ and $2p_0$
 angular velocities in $\phi$, cf. Eqs. \eqref{phi_1s} and \eqref{phi_2p0}, in order to
 assess the numerical results.  For the $1s$ state, Eq. \eqref{phi_1s} becomes
 \begin{equation}
 \label{phi_1s_tau}
 \frac{d \phi}{d \tau} = \frac{\hbar}{ma^2 \omega_0 \xi} = \frac{8}{3 \xi}.
 \end{equation}
 The angular velocity for the $2p_0$ state is simply one-half the above result.
 
\begin{figure}
 \begin{center}
 \epsfig{file=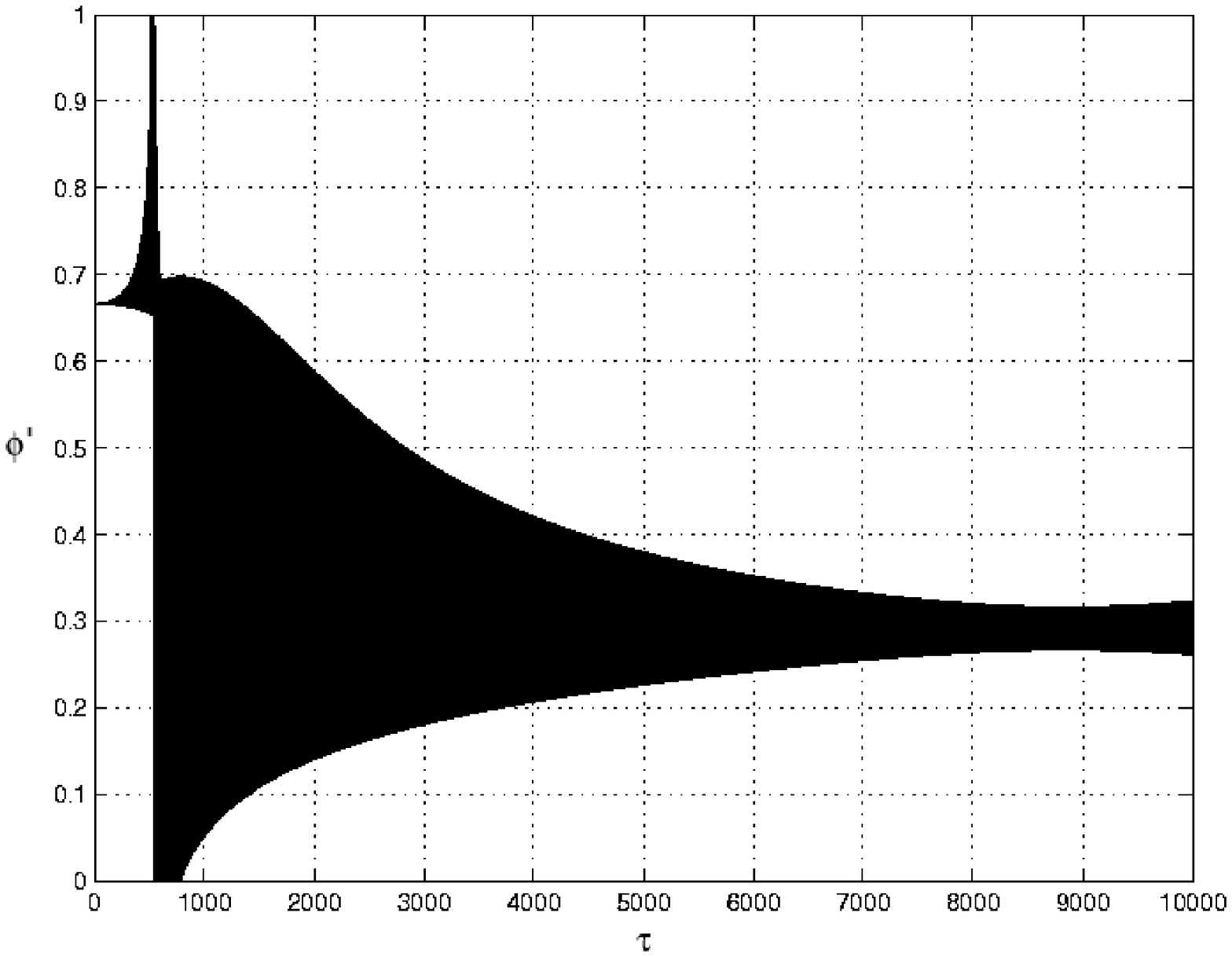,width=10cm}
 \vspace{0.5cm}
 \end{center}
 \caption{Scaled angular velocity $\frac{d\phi}{d\tau}$ along the trajectory shown in Fig. 1}\label{fig:phichangea}
 \end{figure}

\begin{figure}
 \begin{center}
 \epsfig{file=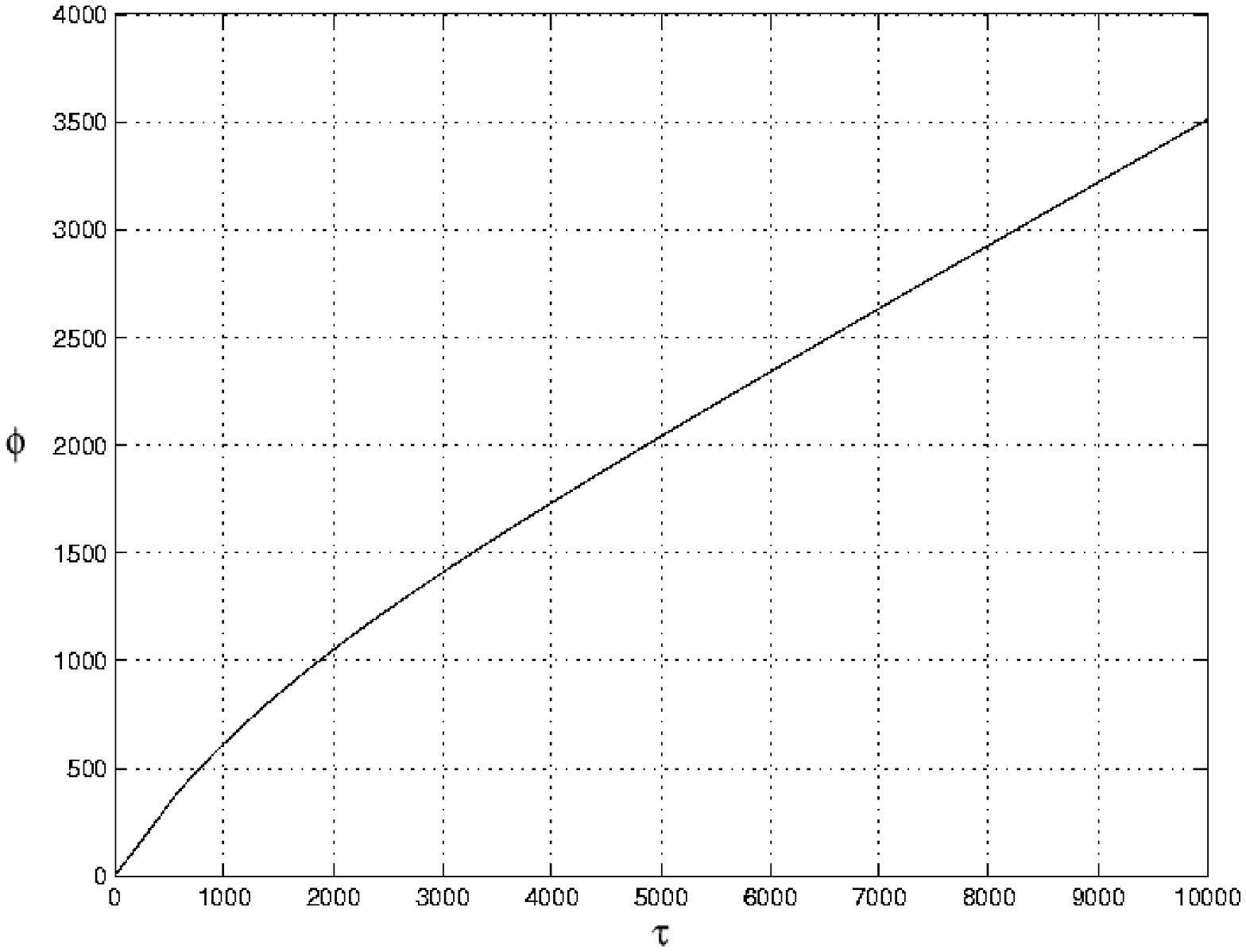,width=10cm}
 \vspace{0.5cm}
 \end{center}
 \caption{Angle $\phi(\tau)$ along the trajectory shown in Fig. 1}\label{fig:phia}
 \end{figure}

 The values of the scaled angular velocity $d \phi / d \tau$ associated with the
 trajectory shown in Fig. \eqref{fig:traja} are presented as a function
 of $\tau$ in Fig. \eqref{fig:phichangea}.
 (In Fig. \eqref{fig:phia} are shown the
 corresponding values of $\phi$.)
 Recall that this trajectory began with the value $\xi(0)=4$.
 >From Fig. \eqref{fig:traja},
 $d \phi / d \tau$ is observed to be roughly $2/3$ near $\tau = 0$.
 This is in agreement with Eq. \eqref{phi_1s_tau} -- the electron is
 revolving about the $z$-axis at  the $1s$ rate.
 
 Near $\tau=9000$, roughly the time for $|c_b(\tau)|^2$ to
 achieve its maximum value, we observe that $d \phi / d \tau \approx 0.3$.
 At that time, $\xi \approx 4.5$.  This is in agreement with
 the scaled $2p_0$ rate $4/(3\xi) \approx 0.3$.
 
 
 In the causal interpretation it is also possible to ascribe a value of
 energy along a trajectory and thereby examine the time-dependent
 behaviour of the energy during a transition. Note that this is in
 contrast to standard quantum mechanics, in which we can only compute
 probabilities of measuring  energy eigenstates  $E_1$ or $E_2$ after
 the perturbation has been turned off.
 One method of computing the energy is to use the result
 $E = -\partial S / \partial t$ implied by the Hamilton-Jacobi
 equation \eqref{sequation}.  However, even when applied to the relatively
 simple wavefunction $\psi$ in Eq. \eqref{tdpsi},
 this method is quite cumbersome.
 On the other hand, Holland's method of assigning values of an observable
 quantity $A$ to points on a trajectory, which we outline below
 (see \cite{Ho93}, pp. 91-93),
 is quite easily implemented in our problem.
 
\begin{figure}
 \begin{center}
 \epsfig{file=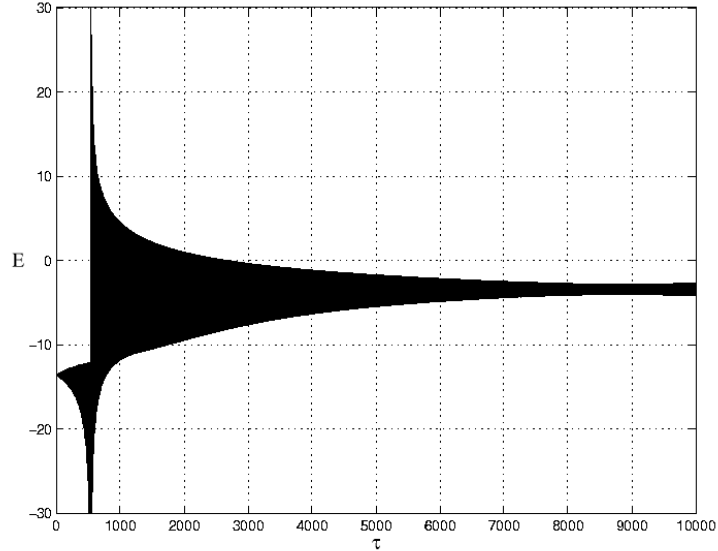,width=10cm}
 \vspace{0.5cm}
 \end{center}
 \caption{Energy (eV) along the trajectory in Fig. \eqref{fig:traja}
 }\label{fig:ena}
 \end{figure}

 The usual expectation value of an observable $\hat{A}$ is
 \[ < \hat{A} > = \bra{\psi}\hat{A}\ket{\psi} =
 \frac{\int\psi^*\hat{A}\psi d^3{\bf x}}{\int\psi^*\psi d^3{\bf x}}.\]
 Since $\hat{A}$ must be Hermitian, ultimately only the real portion of
 the expression contributes to the integral. Hence define the {\it local}
 expectation value to be
 \begin{equation}
  A({\bf x},t) = \frac{\text{Re}\{\psi^*\hat{A}\psi \}}
 {\psi^*\psi}. \end{equation} To find the local energy expectation
 value along the trajectory, we compute this quantity using the
 hamiltonian $\hat{H}=\hat{H}_0+\hat{H}'$ where, as before, $\hat{H}_0$
 is the usual hydrogen hamiltonian, and \[\hat{H}'= -\frac{1}{2}eE_0
 e^{-i\omega t}(r\cos\theta).\] Thus
 \begin{equation}\label{eterms}
  E({\bf x},t)= \frac{\text{Re}\{\psi^*(\hat{H}_0+\hat{H}')\psi
 \}}{\psi^*\psi} =\frac{\text{Re}\{\psi^*\hat{H}_0\psi \}}{\psi^*\psi}+
 \frac{\text{Re}\{\psi^*\hat{H}'\psi \}}{\psi^*\psi}, \end{equation}
 where $\psi({\bf x},t)$ is given by Eq. \eqref{psicombo}. Note that
 the relationship to the classical energy is not direct; this local
 energy expectation value is not a first integral of motion, simply a
 quantity whose average, over trajectories, yields the quantum
 mechanical energy expectation value. Here we will refer to it as the
 local energy.
 After some manipulations, the first term in Eq. \eqref{eterms} 
 is given by
 \[ \frac{\text{Re}\{\psi^*(\hat{H}_0)\psi
 \}}{\psi^*\psi}=  \frac{|c_a|^2E_1\psi_{100}^2 +
 |c_b|^2E_2\psi_{210}^2+\psi_{100}\psi_{210}(E_2+E_1)
 \text{Re}\{c_a c_b e^{-i\omega t}\}}{\psi^*\psi}
 \]
 and the second term is simply
 \[ \frac{\text{Re}\{\psi^*\hat{H}'\psi \}}{\psi^*\psi}=
 \frac{\text{Re}\{\psi^*(-\frac{1}{2}eE_0
 e^{-i\omega t}(r\cos\theta)\psi)\}}{\psi^*\psi}= -\frac{1}{2}eE_0r\cos\theta
 \cos\omega t\]
 where, referring to Eq. \eqref{Tprime},
 \[ \text{Re}\{c_a c_b e^{-i\omega t}\}=\frac{\nu}{2\sigma}(\frac{\Omega}{\sigma}\cos\omega_0 t-\frac{\Omega}{\sigma} \cos\sigma t \cos\omega_0 t -
 \sin\sigma t\sin\omega_o t)
 =T'(t).\]
 Therefore we have
 \begin{equation}
 E = \frac{|c_a|^2E_1\psi_{100}^2 +
 |c_b|^2E_2\psi_{210}^2+\psi_{100}\psi_{210}(E_2+E_1)T'(t)}
 {\psi^*\psi}
 -\frac{1}{2}eE_0r\cos\theta \cos\omega t.
 \end{equation}
 This function can be evaluated along the trajectories shown in
 Figs. \eqref{fig:traja} and \eqref{fig:trajb}. The results for the first
 trajectory are shown in Fig. \eqref{fig:ena}.
 
\begin{figure}
 \begin{center}
 \epsfig{file=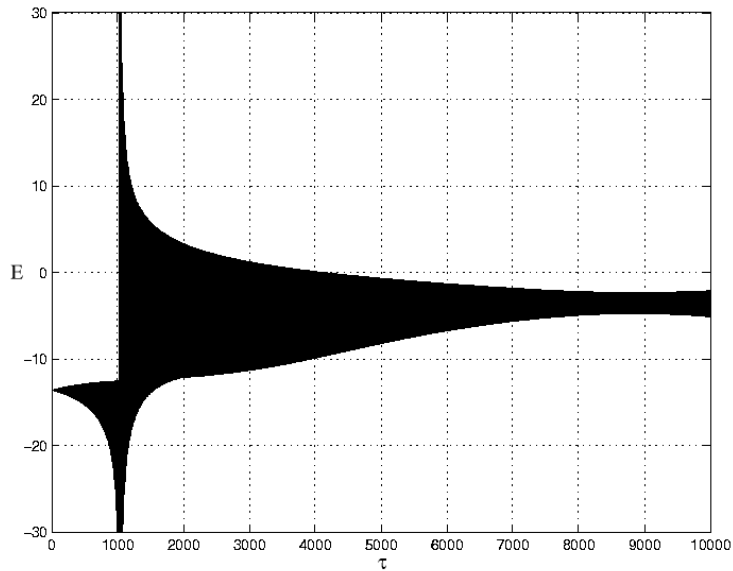,width=10cm}
 \vspace{0.5cm}
 \end{center}
 \caption{Energy (eV) along the trajectory in Fig. \eqref{fig:trajb}
 }\label{fig:enb}
 \end{figure}

 Note that at $\tau=0$, the local energy is equal to the $1s$ ground state energy
 $-13.6 ~ \text{eV}$, and that after the transition, the energy
 oscillates near the value of $E_2=-3.4 ~ \text{eV}$, corresponding to the energy
 of hydrogenic $n=2$ states.   The energy is not constant
 even in a proximity of these points because of
 the extra oscillating perturbation $\hat{H}'$ that represents the semi-classical
 radiation field. The spike in the energy corresponds to a point along the
 trajectory that approaches a zero of the wavefunction.  Intuitively, one can
 understand the appearance of such a spike because the electron 
cannot spend time in regions
 where $\psi^*\psi$ is very small -- it receives a `kick' from the
 quantum potential which corresponds to a sudden rise in its energy.  Plots
 of the quantum potential  for several systems and a discussion are given in
 Holland \cite{Ho93} and references within, for example, \cite{DeHi,PhDeHi}.
 The energy along the trajectory shown in Fig. \eqref{fig:trajb} is shown in
 Fig. \eqref{fig:enb}.
 
 Finally, we mention that we have computed trajectories to higher times.
 At $\tau \approx 18000$, the scaled angular velocity $d \phi / d \tau$
 is observed to oscillate about the $1s$ value.
 The energy is also observed to oscillate about the $1s$ value of -13.6 eV.
 This is in accordance with $|c_b(\tau)|^2 \approx 0$ from Eq. \eqref{cb}
 and the values of the parameters used in the calculation.
 We conclude that the transition has reversed and that the electron
 has returned to the $1s$ ground state.
 
 \section{Concluding remarks}
 
 We have studied
 the problem of a simple $1s$ to $2p_0$ electronic
 transition in hydrogen -- induced by an oscillating (semiclassical)
 radiation field -- in terms of the causal interpretation of quantum
 mechanics.  In contrast to Bohm's original formulation
 ${\bf p} = \nabla S$, however, we have employed an additional
 spin-dependent term $\nabla \log \rho \times {\bf s}$ in the
 equation of motion, where $\rho = \psi^* \psi$.
 The electron is assumed to be in a
 ``spin up'' eigenstate, with associated spin vector
 ${\bf s} = \frac{\hbar}{2} {\bf k}$,
 during the transition.
 The electronic trajectories lie on invariant hyperboloid
 surfaces of revolution about the $z$-axis.  The nature
 of the trajectories over these surfaces is quite complex due to the
 quasiperiodic nature of the equations of motion which, in turn, arise
 from the interplay of the $1s$-$2p$ transition frequency $\omega_0$,
 the frequency $\omega$ of the oscillating radiation and
 $E_0$, the field strength.
 
 As the electron moves over the invariant surface
 it also revolves about the $z$-axis due
 to the spin-dependent momentum term.
 The progress of the transition can be tracked by observing
 the local energy $E$ and the angular velocity $\dot \phi$ of the electron.
 Beginning at values associated with the $1s$ ground state,
 both quantities are seen to evolve toward values associated with
 the $2p_0$ excited state in time.
 The causal interpretation has offered a way
 to examine the phenomenon of transition which is not limited to
 computing the probabilities of measuring $E_1$ and $E_2$ at various
 times after the perturbation has been turned off.
 
 \section*{Acknowledgments}
 We gratefully acknowledge that this
 research has been supported by the Natural Sciences and Engineering
 Research Council of Canada (NSERC) in the form of a Postgraduate
 Scholarship (CC) and an Individual Research Grant
 (ERV).  CC also acknowledges partial financial support from the
 Province of Ontario (Graduate Scholarship) as well as the
 Faculty of Mathematics, University of Waterloo.

 \end{document}